 \documentclass[
a4paper,showpacs,preprint, amsmath, amssymb, floatfix,
nofootinbib,wrapfloat byrevtex]{revtex4}

\usepackage{lineno,hyperref}
\modulolinenumbers[5]

 \usepackage{amsmath,mathrsfs,graphicx,epstopdf,placeins,float}
 \usepackage{booktabs}
 \usepackage{multirow}
 \usepackage[titletoc]{appendix}
 \usepackage{titlesec}
\usepackage{changepage}

 \newcommand{\beq}[1]{\begin{equation}\label{#1}}
 \newcommand{\eeq}{\end{equation}}
 \newcommand{\bea}[1]{\begin{eqnarray}\label{#1}}
 \newcommand{\eea}{\end{eqnarray}}
 \newcommand\figcaption{\def\@captype{figure}\caption}
 \newcommand\tabcaption{\def\@captype{table}\caption}
 \newcommand{\bfk}{\mathbf{k}}

\begin{document}
 \title{Spectrum of supersymmetric and bosonic open 2-branes}
 \author{Muhammad Abdul Wasay}
\email{muhammad.wasay@uaf.edu.pk}
\affiliation{Department of Physics, University of Agriculture\\ Faisalabad 38040, Pakistan}
 \begin{abstract}
    We consider both the supersymmetric open 2-brane and bosonic open
2-brane, their quantization and spectrum under the flat metric condition. The
supersymmetric spectrum turns out to be discrete, while the spectrum of purely
bosonic open 2-brane is shown to be devoid of any massless states.
 \end{abstract}
 \maketitle
 \smallskip

\section{Introduction}
 \vspace{2mm}
 String theory emerged as a candidate for a consistent theory of quantum gravity, however, the route turned out to be more and more subtle by the passage of time. It was the time to think of an alternate and thus the birth of supermembranes\cite{super1,super2,super3} A $p$-brane is a $p$-dimensional extended object which can move freely in a $D$-dimensional space-time, and $D\geq p$ and for a string, we have $p=1$. When a $p$-brane moves in space-time, it sweeps out a $p+1$ dimensional world-volume. It was later shown\cite{super4} that a supermembrane is unstable because it has a continuous spectrum. Most of the trouble with brane quantization is because the membrane is known to be a difficult system to analyze precisely because there is no analogue of the conformal gauge like in string theory. This has been known for a long time and was discussed in detail for the bosonic case by Collins and Tucker\cite{collins}.

There are plenty of works on string quantization compared to brane quantization because of the difficulty which Nambu action meets when one tries to generalize it from $1\!+\!1$ to $2\!+\!1$ dimensions. However, it is still possible to quantize a bosonic membrane in a flat background\cite{huang,huang1}. In this paper we will focus our attention to the case of open 2-brane. We will start with a Polyakov like action for a supersymmetric open 2-brane\cite{mine1} and discuss its spectrum, this will be done in RNS formalism and in $D=10$. We will also consider the spectrum of purely bosonic open 2-brane\cite{mine2} in $D=26$.
\section{The Action and Hamiltonian}

The dynamics of the 2-brane is governed by the Polyakov-like action\cite{mine1} supplemented by a fermionic part

\begin{equation}
S\!=\!\frac{-1}{4\pi\alpha^\prime}\!\int\!\! d^3\sigma\sqrt{-h}[h^{ab}\partial_aX^\mu\partial_bX_\mu+h^{ab}\bar\Psi^\mu\gamma_bD_a\Psi_\mu+R-2\Lambda
\\
-\bar\chi_\mu\gamma^{\mu\nu\rho}D_\nu\chi_\rho]
\label{action}
\end{equation}
where $\mu=0,1,...,9$; $d^3\sigma=d\tau d\sigma^1 d\sigma^2$ and $\chi_rho$ is the graviton. $\gamma^a$ are the 2 dimensional representations of Dirac algebra. We only can consider part of the brane dynamics, i.e., 3 dimensional supergravity coupled to $X$ and $\Psi$ because considering the full dynamics leads to a continuous spectrum\cite{super4}. In addition to this, we will work in a flat background. Imposing these conditions lead to a refined version of the action \eqref{action}, in which the last three terms drop out and covariant derivatives become partial derivatives.  The supersymmetry transformations are given by
\begin{equation}
\delta X^\mu=\bar\epsilon\Psi^\mu
 \qquad \textmd{and}\qquad
 \delta\Psi^\mu=\gamma^a\partial_aX^\mu\epsilon
\end{equation}
Moreover, in a 10-dimensional target space the supersymmetry of 2-brane is only possible when we include the gauge field degree of freedom on the world-volume of the brane.
The energy-momentum tensor is given by
\begin{eqnarray}
T_{ab}\!
=\!\frac{1}{2}[\partial_aX^\mu\partial_bX_\mu\!-\!\frac{1}{2}h_{ab}h^{cd}\partial_cX^\mu\partial_dX_\mu\!
-\!\frac{1}{2}h_{ab}h^{cd}\bar\psi^\mu\gamma_d\partial_c\psi_\mu\!
+\!\frac{1}{2}\bar\psi^\mu\gamma_b\partial_a\psi_\mu\!+\!\frac{1}{2}\bar\psi^\mu\gamma_a\partial_b\psi_\mu ]
\end{eqnarray}
The equation of motion for the bosonic fields $X^\mu$ is
\begin{equation}
(\partial_\tau^2-\partial_1^2-\partial_2^2)X^\mu(\tau,\sigma^1,\sigma^2)=0
\end{equation}
and for the fermionic fields $\Psi^\mu$
\begin{equation}
\gamma^a\partial_a\psi^\mu(\tau,\sigma^1,\sigma^2)=0
\end{equation}
The boundary conditions on fermions are (anti)periodic when we work in (NS)R sectors respectively. For the bosons we impose Neumann boundary conditions so that the ends of the 2-brane are free to move in the space-time.
\section{Modes expansions and commutation/anticommutation relations}
The mode expansions for fermions are integrally (half integrally) moded for R (NS) sector respectively. For the R sector, these are given by
\begin{eqnarray}
\psi^\mu(\sigma)=\frac{1}{(2\pi)^2}\int\limits_{0}^\infty\!\!\!\frac{d^2k}{\sqrt{2\omega}}\big(
d_{\bfk\textbf{s}}^\mu e^{-ik_a\sigma^a}+d_{\bfk\textbf{s}}^{\mu\dagger} e^{ik_a\sigma^a}\big)u_{\bfk\textbf{s}}\\
\bar\psi^\mu(\sigma)=\frac{1}{(2\pi)^2}\int\limits_{0}^\infty\!\!\!\frac{d^2k}{\sqrt{2\omega}}\big(
d_{\bfk\textbf{s}}^{\mu\dagger} e^{ik_a\sigma^a}+d_{\bfk\textbf{s}}^\mu e^{-ik_a\sigma^a}\big)\bar u_{\bfk\textbf{s}}
\end{eqnarray}
where $``d"$ is the R sector oscillator. There is a similar expansion for NS sector fermions, but with $``b"$ oscillators, which will be half integrally moded in the discrete limit\cite{mine1}.
The modes expansion for bosons is given by\cite{huang,huang1,mine1,mine2}
\begin{eqnarray}
X^\mu(\sigma)=\frac{x^\mu}{\sqrt\pi}+\frac{2\acute{\alpha}p^\mu}{\sqrt\pi}\tau+i\sqrt{2\acute{\alpha}}\sum\limits _{m,n=0}^{+\infty}(n^2+m^2)^{-\frac{1}{4}}
~~~~~~~~~~~~~~~~\nonumber\\
\times
\left(X^\mu_{nm}e^{i\tau\sqrt{n^2+m^2}}-X^{\dag \mu}_{nm}e^{-i\tau\sqrt{n^2+m^2}} \right)
\times\textmd{cosn}\sigma^1\textmd{cosm}\sigma^2
\label{bosonic1}
\end{eqnarray}
 The commutation/anticommutation relations for world-volume bosons and fermions are obtained\cite{mine1,mine2} by using the relevant mode expansions.

\section{Hamiltonian and spectrum}
The Hamiltonian of the supersymmetric open 2-brane (R sector) is given by
\bea{}
 &&\hspace{-5mm}\mathcal H\!=\eta_{\mu\nu}\!\sum\limits_{n=1}^\infty n\left(X_{n0}^{\mu\dag}X_{n0}^\nu+\frac{1}{2}\eta^{\mu\nu} \right)+
  \eta_{\mu\nu}\!\sum\limits_{m=1}^\infty m\left(X_{0m}^{\mu\dag}X_{0m}^\nu+\frac{1}{2}\eta^{\mu\nu} \right)\nonumber
  \\
 &&\hspace{-5mm}
 ~+\eta_{\mu\nu}\sum\limits_{n,m=1}^\infty\sqrt{n^2+m^2}\left(X_{nm}^{\mu\dag}X_{nm}^\nu+\frac{1}{2}\eta^{\mu\nu} \right)-\alpha'M^2
 \nonumber \\
 &&\hspace{-5mm}
 ~+\eta_{\mu\nu}\!\sum\limits_{n'=1}^\infty n'\left(d_{n'0}^{\mu\dag}d_{n'0}^{\nu}-\frac{1}{2}\eta^{\mu\nu}\delta_{n'n'} \right)+
 \eta_{\mu\nu}\!\sum\limits_{m'=1}^\infty m'\left(d_{m'0}^{\mu\dag}d_{m'0}^{\nu}-\frac{1}{2}\eta^{\mu\nu}\delta_{m'm'} \right)
 \nonumber \\
 &&\hspace{-5mm}
 ~+\eta_{\mu\nu}\!\!\!\!\!\sum\limits_{n',m'=1}^\infty\!\!\!\!\sqrt{n^{\prime2}+m^{\prime2}}\left(d_{n'm'}^{\mu\dag}d_{n'm'}^{\nu}-\frac{1}{2}\eta^{\mu\nu}\delta_{n'n'}\delta_{m'm'} \right)
  \hspace{-5mm}
  \label{superhamiltonian}
 \eea
 The Hamiltonian for purely bosonic open 2-brane\cite{mine2} is obtained by using the mode expansions, Eqs \eqref{bosonic1} and its canonically conjugate momentum
 \bea{}
4\pi\alpha^\prime H
=2\alpha^\prime\pi^2\eta_{\mu\nu}\sum\limits_{n=1}^\infty n\left[X^{\mu\dag}_{n0}X^\nu_{n0}+X^\mu_{n0}X^{\nu\dag}_{n0}\right]~~~~~~~~~~~~~~~~~~
\label{hamiltonianof2brane}\nonumber\\
+2\alpha^\prime\pi^2\eta_{\mu\nu}\sum\limits_{m=1}^\infty m\left[X^{\mu\dag}_{0m}X^\nu_{0m}+X^\mu_{0m}X^{\nu\dag}_{0m}\right]+4\pi\alpha^{\prime2}p^2
\nonumber\\
~~~~+\alpha^\prime\pi^2\eta_{\mu\nu}\sum\limits_{n,m=1}^\infty(n^2+m^2)^{\frac{1}{2}}\left[X^{\mu\dag}_{nm}X^\nu_{nm}+X^\mu_{nm}X^{\nu\dag}_{nm} \right]
\eea

\subsection{Spectrum of supersymmetric open 2-brane}
For the supersymmetric case, the normal ordering constants in the R sector exactly cancel as a consequence of world-volume supersymmetry, so we get a positive mass formula
\bea{}
\alpha^\prime M^2=\sum\limits_{n=1}^\infty n\eta_{\mu\nu}\left(X_{n0}^{\mu\dag}X_{n0}^\nu+d_{n0}^{\mu\dag}d_{n0}^{\nu} \right)+\sum\limits_{m=1}^\infty m\eta_{\mu\nu}\left(X_{0m}^{\mu\dag}X_{0m}^\nu+d_{0m}^{\mu\dag}d_{0m}^{\nu} \right)+\nonumber
\\
\sum\limits_{n,m=1}^\infty\sqrt{n^2+m^2}\eta_{\mu\nu}\!\left(X_{nm}^{\mu\dag}X_{nm}^\nu\!+d_{nm}^{\mu\dag}d_{nm}^{\nu}\right)~~~~~~~~~~~~~~\label{Massformulasusic}
\\
\Rightarrow\alpha^\prime M^2=N_n+N_m+N_{nm}~~~~~~~~~~~~~~~~~~~~~~~~~~~~~~~~~~~~~~~~~~~~~~~~
\eea
with $N_n$, $N_m$ and $N_{nm}$ being the first, second and third sum in Eq.\eqref{Massformulasusic}. For the NS sector, where the fermionic operators $b_{nm}$, $b_{0m}$ and $b_{n0}$ are half integrally moded, the mass formula reads
\bea{}
\alpha^\prime M^2=N_n+N_m+N_{nm}+a+b
\eea
with $a$ and $b$ being the normal ordering constants
\bea{}
a=\eta^\mu_\mu\sum\limits_{n=1}^\infty n -\eta^\mu_\mu\sum\limits_{n=\frac{1}{2}}^\infty n \delta_{nn} ~~~~~~~~~~~~~~~~~~~~~~~~~~~~~~~~~~~~\\ b=\frac{1}{2}\!\sum\limits_{n,m=1}^\infty\!\sqrt{n^2+m^2}\eta^\mu_\mu\!-\!\frac{1}{2}\!\!\sum\limits_{n,m=\frac{1}{2}}^\infty\!\!\sqrt{n^2+m^2}\eta^\mu_\mu\delta_{nn}\delta_{mm}
\eea
A Zeta function regularization of $a$ gives $a=-2$ and an Epstein Zeta function regularization of $b$ should give $b=\frac{3}{2}$, in order to get a supersymmetric spectrum. The spectrum of states for R and NS sectors are displayed in the tables below\cite{mine1}.

\renewcommand{\arraystretch}{1.5}
\begin{tabular}{|c|c|l|c|c|}
\multicolumn{4}{c}{\textbf{Table 1: Spectrum of states for R-sector.}}\\
\hline
\multicolumn{1}{|c|}{$\alpha^\prime M^2$}
&\multicolumn{1}{|c|}{States}
&\multicolumn{1}{|c|}{$\gamma_{11}$}
&\multicolumn{1}{|c|}{Little Group}
&\multicolumn{1}{|c|}{Representation}\\
\hline
  \multirow{2}{*}0 & $|\zeta\rangle$ & $+1$ & \multirow{2}{*}{SO(8)} & \multirow{2}{*}{$8\oplus8$} \\
   
   & $|\bar\zeta\rangle$ & $-1$ &  &\\
   \hline
   \multirow{4}{*}{+1} & $\alpha_{-1,0}^{i}|\zeta\rangle ;d_{-1,0}^{i}|\bar\zeta\rangle;$ &$+1$ & \multirow{4}{*}{SO(9)}& \multirow{4}{*}{512}\\
      & $\alpha_{0,-1}^{i}|\zeta\rangle;d_{0,-1}^{i}|\bar\zeta\rangle;$ &+1 & &\\
      & $\alpha_{-1,0}^{i}|\bar\zeta\rangle;d_{-1,0}^{i}|\zeta\rangle;$ &$-1$ & &\\
      & $\alpha_{0,-1}^{i}|\bar\zeta\rangle;d_{0,-1}^{i}|\zeta\rangle$ &$-1$ & & \\
   \hline
  \end{tabular}\\
\hspace{15mm}
  \renewcommand{\arraystretch}{1.5}
\hskip-1.8cm\begin{tabular}{|c|c|l|c|c|}
\multicolumn{4}{c}{\textbf{Table 2: Spectrum of states for NS-sector.}}\\
\hline
\multicolumn{1}{|c|}{$\alpha^\prime M^2$}
&\multicolumn{1}{|c|}{States}
&\multicolumn{1}{|c|}{$G$}
&\multicolumn{1}{|c|}{Little Group}
&\multicolumn{1}{|c|}{Representation}\\
\hline
  $-\frac{1}{2}$ & $|0\rangle$ & $-1$ & SO(9) & $1$ \\
   \hline
   0 & $b^i_{\!-\frac{1}{2},0} ;b^i_{0,-\frac{1}{2}}$ & $+1$ & SO(8) &$8\oplus8$\\
   \hline
   \multirow{2}{*}{$+\frac{1}{2}$} & $\alpha^i_{-1,0};\alpha^i_{0,-1}$& \multirow{2}{*}{$-1$}&  \multirow{2}{*}{SO(9)}& \multirow{2}{*}{$8\oplus8\oplus64$}\\
      & $b^i_{\!-\frac{1}{2},0}b^j_{\!-\frac{1}{2},0};b^i_{0,-\frac{1}{2}}b^j_{0,-\frac{1}{2}};$ $b^i_{\!-\frac{1}{2},0}b^j_{0,\!-\frac{1}{2}}$ & & &\\
     \hline
     \multirow{5}{*}{$+1$}   & $\alpha^i_{-1,0}b^j_{\!-\frac{1}{2},0};\alpha^i_{0,-1}b^j_{\!-\frac{1}{2},0};$ &\multirow{5}{*}{$+1$} &\multirow{5}{*}{SO(9)} &\multirow{5}{*}{512}\\
     & $\alpha^i_{-1,0}b^j_{0,-\frac{1}{2}};\alpha^i_{0,-1}b^j_{0,-\frac{1}{2}};$& & &\\
     & $b^i_{\!-\frac{1}{2},0}b^j_{\!-\frac{1}{2},0}b^k_{\!-\frac{1}{2},0};$ & & &\\
     & $b^i_{0,-\frac{1}{2}}b^j_{0,-\frac{1}{2}}b^k_{0,-\frac{1}{2}};$ 
      $b^i_{-\frac{1}{2},0}b^j_{0,-\frac{1}{2}}b^k_{-\frac{1}{2},0};$ & & &\\
     & $b^i_{-\frac{1}{2},0}b^j_{0,-\frac{1}{2}}b^k_{0,-\frac{1}{2}};$ 
    $b^{i'\pm}_{-\frac{3}{2},0};b^{i'\pm}_{0,-\frac{3}{2}}$ & & & \\
   \hline
  \end{tabular}\\
  The states $b^{i'\pm}_{-\frac{3}{2},0}$ are coming from the gauge field introduced into the world-volume of the 2-brane. After truncation by a GSO-like condition, the above spectrum is supersymmetric.
\subsection{Spectrum of bosonic open 2-brane}

 The purely bosonic theory is consistent in $D=26$. The Hamiltonian in Eq. \eqref{hamiltonianof2brane} can be written as
 \bea{}
 H=N_1+N_2+N_{12}+a+b-\alpha^\prime M^2
 \eea
 where $N_1$, $N_2$ and $N_{12}$ contain purely bosonic variables and $a$, $b$ arise from the normal ordering. Using Epstein Zeta functions, the lowest three mass levels for a bosonic open 2-brane are
 \bea{}
 \alpha^\prime M^2&=&-\frac{3}{2}
 \\
 \alpha^\prime M^2&=&-\frac{1}{2}
 \\
 \alpha^\prime M^2&=&\frac{1}{2}
 \eea

 The spectrum is devoid of massless states\cite{mine2}.
 
 \section{Summary and conclusion}
We studied the spectrum of supersymmetric and bosonic open 2-brane, in $D=10$ and $D=26$, respectively. We showed that partly fixing the phase space of 2-brane dynamics can lead to a supersymmetric and discrete spectrum of states. It is also shown in the bosonic 2-brane case that the spectrum does not contain any massless states to play the role of gravitons, and the spectrum only contains half integer mass squared values.

\end{document}